\documentclass[aps,pre,twocolumn,superscriptaddress,showpacs,nofootinbib]{revtex4}
\usepackage[T1]{fontenc}
\usepackage[latin9]{inputenc}
\usepackage{amsmath}
\usepackage{graphicx}
\usepackage{amssymb}
\usepackage{esint}
\usepackage{nicefrac}
\usepackage{dsfont}
\usepackage[normalem]{ulem}
\usepackage[usenames]{color}

\begin{document}
\title{Actively stressed marginal networks}

\author{M. Sheinman}
\affiliation{Department of Physics and Astronomy, VU University, Amsterdam, The Netherlands}

\author{C. P. Broedersz}
\affiliation{Lewis-Sigler Institute for Integrative Genomics and the Department of Physics, Princeton University, Princeton, NJ 08544, USA}

\author{F. C. MacKintosh}
\affiliation{Department of Physics and Astronomy, VU University, Amsterdam, The Netherlands}

\date{\today}

\begin{abstract}
We study the effects of motor-generated stresses in disordered three dimensional fiber networks using a combination of a mean-field, effective medium theory, scaling analysis and a computational model. We find that motor activity controls the elasticity in an anomalous fashion close to the point of marginal stability by coupling to critical network fluctuations. We also show that motor stresses can stabilize initially floppy networks, extending the range of critical behavior to a broad regime of network connectivities below the marginal point. Away from this regime, or at high stress, motors give rise to a linear increase in stiffness with stress. Finally, we demonstrate that our results are captured by a simple, constitutive scaling relation highlighting the important role of non-affine strain fluctuations as a susceptibility to motor stress. 
\end{abstract}

\maketitle

The mechanical properties of cells are regulated in part by internal stresses generated actively by molecular motors in the cytoskeletal filamentous actin network \cite{Howard2001Mechanics}. On a larger scale, collective motor activity allows the cell to contract the surrounding extracellular matrix, consisting also of biopolymer networks. Experiments show that such active contractility dramatically affects network elasticity, both in reconstituted intracellular F-actin networks with myosin motors \cite{Mizuno2007Nonequilibrium,Bendix2008Quantitative,Koenderink2009Active,Gordon2012Hierarchical} and in extracellular matrices with contractile cells \cite{Lam2011Mechanics}.  The dynamics and elasticity of active biopolymer networks have been studied theoretically using long-wavelength hydrodynamic approaches as well as affine models \cite{MacKintosh2008Nonequilibrium,Liverpool2009Mechanical,Shokef2012Scaling}. These approaches, however, fail to describe highly disordered networks. There is also experimental evidence that cytoskeletal networks may be unstable or only marginally stable in the absence of motor activity \cite{Cai2010Cytoskeletal}. In such cases, networks are expected to be governed by highly nonuniform, soft or floppy modes of deformation that may lead to a fundamental breakdown or failure of continuum elasticity \cite{Broedersz2011Criticality}. Importantly, motor-induced contractile stresses can be expected to couple to these soft modes, giving rise to a nonlinear elastic response that is distinct from the nonlinearities arising from single fiber elasticity that have been considered in previous models. Moreover, such a coupling to local soft modes of the network may call into question the equivalence of internal (motor) and external stress, a tacit assumption in the analysis of recent in vitro experiments \cite{Mizuno2007Nonequilibrium, Koenderink2009Active}.

Here, we introduce a simple model to study the effects of motor generated stresses in disordered fiber networks. Networks are formed by crosslinked straight fibers with linear stretching and bending elasticity. These fibers are organized on a face centered cubic (FCC) lattice in which a certain fraction of the the bonds can randomly be removed. Motor activity is introduced by contractile force dipoles acting between neighboring network nodes. We find that motors can stabilize the elastic response of otherwise floppy, unstable networks. The motor stress also controls the mechanics of stable networks above a characteristic threshold, in the vicinity of which the network exhibits critical strain fluctuations. We develop a quantitative effective medium theory to describe the elastic response of these systems. Interestingly, the network's stiffness is controlled by a coupling of the motor induced stresses to the strain fluctuations. This coupling gives rise to anomalous regimes at the stability thresholds, at which network criticality is reflected in both divergent strain fluctuations and anomalous dependences of the network mechanics on stress. In these critical regimes, the shear modulus depends nonlinearly on both motor stress and single filament elasticity
\cite{Lam2011Mechanics, Ehrlicher2011Cell, Broedersz2011Molecular, Chen2011Strain}. Interestingly, this dependence on internal motor stress differs qualitatively from that of an applied external stress. 

A key parameter that characterizes fiber networks is the mean coordination number, $z$. Although the network is connected above a threshold $z=z_{\rm cond}\simeq 2$, it only becomes rigid above a higher rigidity threshold $z_b\simeq 3.4$. This threshold is due to the bending rigidity of the individual fibers and it lies below the central-force (CF) rigidity threshold, $z_{\rm cf}\simeq 6$, for a spring-only network. In general, when some fraction of the bonds are under stress, additional \emph{constraints} are introduced \cite{Huisman2011Internal}. More formally, these constraints appear as scalar terms in the Hamiltonian \cite{Alexander1998}. These additional stress-constraints may shift the various rigidity thresholds in the system. In random spring networks, for example, this can be realized by applying finite network deformations; this has been studied in spring networks \cite{tang1988percolation,wyart2008,Sheinman2012Nonlinear} where the actual rigidity threshold shifts continuously to lower values with the applied external strain. Under such external deformations, the internal stress is free to adopt the most favorable distribution. By contrast however, motors impose a fixed distribution of internal stress, which may lead to a qualitatively different network mechanics.

To provide insight into the elasticity of fibrous networks with contractile \emph{internal} stresses, we use a model of fibers organized on a FCC lattice. By removing lattice-bonds with a probability $1-p$, we tune the average coordination number, $z=\mathcal{Z}p$, where $ \mathcal{Z}=12 $ for the undiluted lattice. Motors are introduced as contractile force dipoles and are inserted randomly with a probability $q$. The fibers are modeled as linear elastic beams with a stretching modulus $\mu$ and bending rigidity $\kappa$. Using units in which $\ell_0=\mu=1$, the total energy can be written as
\begin{align}
H&=\frac{1}{2}\sum\limits_{\langle ij \rangle}P_{ij} 
\left( \lvert  \mathbf{r}_{ij} \rvert-1 \right)^2  \nonumber\\ 
&+\frac{\kappa}{2}\sum\limits_{\langle ijk \rangle}P_{ij}P_{jk} 
 \left( \frac{ \mathbf{r}_{ij} \times  \mathbf{r}_{jk}}{\lvert  \mathbf{r}_{ij} \rvert \lvert  \mathbf{r}_{jk} \rvert}  \right)^2 \nonumber\\ 
&+f\sum\limits_{\langle ij \rangle} Q_{ij} \lvert  \mathbf{r}_{ij} \rvert
\label{Hamiltonian}
\end{align}
where, $ \mathbf{r}_{ij}=\mathbf{r}_{i}-\mathbf{r}_{j} $ and $ \mathbf{r}_{i} $ denotes the position of $ i$'th node and  $P_{ij}=1$ for present bonds or $P_{ij}=0$ for removed bonds. The first sum extends over neighboring pairs of vertices. The crosslinks themselves do not contribute a torsional stiffness and, thus, the second sum only extends over \emph{coaxial} nearest neighbor bonds on the same fiber. The last term represents the work performed by the motors, where $Q_{ij}=1$ if a motor acts between nodes $i$ and $j$ and $Q_{ij}=0$ otherwise.

To develop a mean-field, effective medium theory (EMT) that captures the disordered nature of  this model---including internal stresses---we extend the theory for the linear mechanical response of disordered spring networks ~\cite{Feng85,Schwartz1985Behavior,Mao2010Soft}. In our EMT approach we ignore the bending contribution ($ \kappa=0$), allowing us to circumvent the difficulties involved in an EMT with three-point bending interactions~\cite{Broedersz2011Criticality,Das2007Effective,Mao2011Effective}. 
Our EMT is based on a mapping between the disordered and an ordered network with an effective elastic constant, yet with the same underlying lattice geometry and under the same internal stress as the original disordered system. The effective elastic constant, $\widetilde{\mu}\left(\sigma_M\right)$, is determined by a self-consistency condition; the local distortion in the effective medium induced by replacing a bond, selected randomly from the disordered system, should vanish on average. For a general disordered network this procedure yields an implicit expression for the effective stretch modulus (see Appendix)
\begin{equation}
\int_{0}^{\infty}\frac{\mu_{ij}-\widetilde{\mu}\left(\sigma_M\right)}{\mu_{EM}+\mu_{ij}-\widetilde{\mu}\left(\sigma_M\right)} \mathcal{P}\left(\mu_{ij}\right)d\mu_{ij}=0,
\label{eq:mEff_Integral0}
\end{equation}
where $ \mu_{EM} $ is the displacement of a bond in the unperturbed effective medium due to a unit force acting along the bond,
 $\mu_{ij}$ is the stretching modulus between nodes $i$ and $j$ and $\mathcal{P}\left(\mu_{ij}\right)$ is the probability density of the moduli in the disordered system. For the case of a  diluted lattice considered here, $ \mathcal{P}\left(\mu_{ij}\right)=p \delta\left(\mu_{ij}-1\right)+
\left( 1-p \right) \delta\left(\mu_{ij}\right) $, we find the EMT shear modulus
\begin{equation}
G_{EM}= \frac{5\sqrt{2}}{72}\widetilde{\mu}\left(\sigma_M\right)+\frac{5}{6}\sigma_M,
\end{equation}
where, within the EMT, $\sigma_M=\sqrt{8} q f$.

While the full expression for $\widetilde{\mu}\left(\sigma_M\right)$ is long (see Appendix), the scaling predictions of the EMT are simple. 
Even below the central-force isostatic point, $z_{\rm cf}$, motor activity induces a finite shear modulus. Far from $z_{\rm cf}$, $G \sim G_0+\sigma_M$, where $G_0$ is the shear modulus of the unstressed network\footnote{In all scaling relationships with additive contributions, unknown numerical prefactors are omitted.}. By contrast, close to $z_{\rm cf}$ there is an anomalous scaling regime $G \sim \sigma_M^{1/2} \mu^{1/2}$.

\begin{figure}[tp]
\includegraphics[width=\columnwidth]{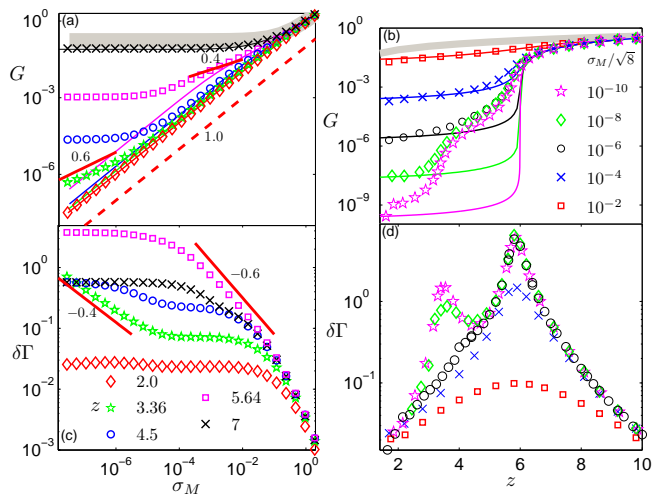}
\caption{(a) Shear modulus is shown vs. motor stress for different values of the coordination number $ z $. Markers represent the numerical data (see legend in (c)). Thin, solid lines represent the results of the EMT for the same values of $ z $ while upper lines correspond to higher values of $ z $. Lines with numbers represent the corresponding slopes and thick grey area contains the affine predictions for all presented coordination numbers. (b) $ G $ is shown vs. $ z $ for different values of $ \sigma_M $. Markers represent numerical data (see legend). Thin, solid lines represent the results of the EMT for the same values of $ \sigma_M $ while upper lines correspond to higher values of $ \sigma_M $. Thick grey area contains the affine predictions for all presented values of $ \sigma_M $. (c) The differential non-affinity parameter is presented vs. motor stress for different values of $ z $ (see legend). (d) $ \delta\Gamma $ is presented vs. $ z $ for different values of $ \sigma_M $ (see legend in (b)). For all data sets $ \kappa=10^{-5} $.}
\label{Fig1}
\end{figure}

\begin{figure}[tp]
\includegraphics[width=\columnwidth]{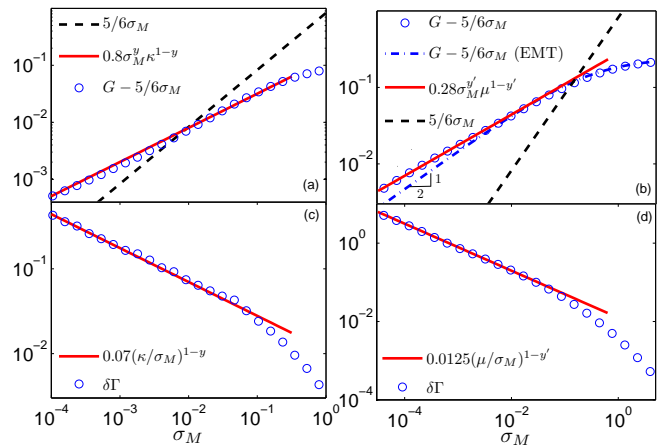}
\caption{Demonstration of the anomalous regimes. (a) Shear modulus separated to the linear part, $ 5/6\sigma_M $ (dashed line), and the rest, $ G-5/6\sigma_M $ (circles) is shown for $ z=3.35 $ (close to the bending rigidity percolation point, $ z_b \simeq 3.36 $) and $ \kappa=10^{-2} $. In the small stress limit the $ G-5/6\sigma_M $ part is found to scale as $\sigma_M^{1-y} $ with $ y \simeq 0.6 $ (solid line). (b) The same analysis as in (a) is shown for $ z=5.9 $ (close to the central-force rigidity percolation point, $ z_{\rm cf} \simeq 5.64 $) and $ \kappa=0 $. In the small stress limit the $ G-5/6\sigma_M $ part is found to scale as $\sigma_M^{1-y'} $ with $ y' \simeq 0.4 $ (solid line), in contrast to the mean-field prediction (indicated by the dashed-dotted line) that scales as $\sqrt{\sigma_M}$ in the small stress limit. (c) The value of $ \delta \Gamma $ (circles) is presented for the same set of parameters as in (a). In the small stress limit it scales as $\sigma_M^{-y} $ (solid line). (d) The value of $ \delta \Gamma $ (circles) is presented for the same set of parameters as in (b). In the small stress limit it scales as $\sigma_M^{-y'} $ (solid line).}
\label{Fig2}
\end{figure}

To test the implications of the EMT, we perform simulations of fiber networks with finite bending rigidities. The shear modulus, $G$, is determined numerically by applying a shear strain along the $111$-plane using Lees-Edwards periodic boundary conditions and energy minimizations are performed by a conjugate gradient algorithm \cite{NumericalRecipesInC}. 
First, we consider the high motor density limit $q\simeq1$. The EMT prediction is in good quantitative agreement with the numerical results over a broad range of network connectivity and motor stress, as shown in Fig. \ref{Fig1} a,b. Since we neglected the contribution of fiber bending energies in the EMT, it fails in the regime where $G$ is governed by $ \kappa$. In addition, in the vicinity of $z_{cf}$ for $ \sigma_M \gg\kappa$, we find a mixed regime, $ G \sim \mu^{1-y'}\sigma_M^{y'} $, where $y'=0.4$, whereas in the EMT $y'=0.5$ (Fig. \ref{Fig2}). This mixed regime is similar in nature to the $\kappa$-$\mu$ coupled mechanical regime around $z_{cf}$ in unstressed fibrous networks \cite{Broedersz2011Criticality}. More generally, such coupled regimes arise in the vicinity of a stability threshold, when there are additional interactions or fields that stabilize the network below the threshold \cite{wyart2008}. Thus, in this model the motor stress acts as an external field. In fact, as may be expected, another anomalous regime is observed in the simulations at the bending rigidity threshold, $ G \sim \kappa^{1-y}\sigma_M^y $, with $y=0.6$ (Figs. \ref{Fig1} and \ref{Fig2}). 

We gain additional physical insight into the elastic properties of active networks with a scaling argument we estimate the amount of work that is performed by the motors when the system is sheared. The characteristic deformation of a single bond in such a network will be such that it avoids energetically costly stretching contributions. Such deformations are oriented perpendicularly to the direction of the bond: the \emph{nonaffine} contribution to this deformation can be estimated by  $\delta u_\perp \sim \gamma \sqrt{\delta \Gamma}$, where the differential nonaffinity parameter is defined as,
\begin{equation}
\delta\Gamma=\frac{1}{\gamma^2}\left\langle \left( \delta \textbf{u}_k-\delta \textbf{u}_k^{\rm aff}\right)^2 \right\rangle_k.
\end{equation}
Here $\delta \textbf{u}_k $ is the displacement of node $k$ under an infinitesimal external shear $\gamma$, $ \delta \textbf{u}_k^{\rm aff} $ is the affine prediction and the average is taken over all network nodes. Interestingly however, this is not the only relevant contribution to the deformation of the bond.
The component of the \emph{affine} deformation perpendicular to the bond does not contribute to bond-stretching energies to harmonic order and, thus, is not avoided. Importantly however, this deformation does contribute to the motor work. Therefore, the total work performed by the internal stress resulting from such deformations scales as $\delta W\sim \sigma_M \gamma^2 \delta\Gamma+\sigma_M \gamma^2$, implying the following relationship for the shear modulus, 
\begin{equation}\label{Gscaling}
G\sim G_0+ \sigma_M \delta\Gamma+\sigma_M.
\end{equation}

\begin{figure}[tp]
\includegraphics[width=\columnwidth]{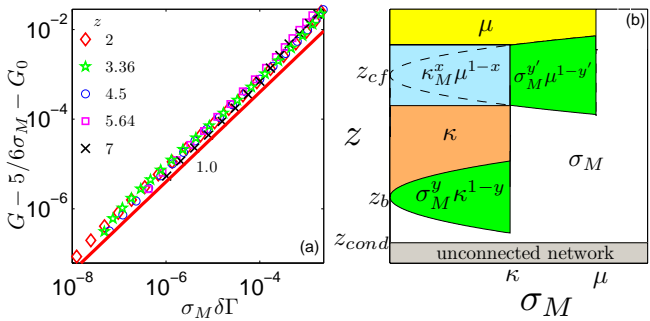}
\caption{(a) Collapse on the data presented on Fig. \ref{Fig1}(a) based on Eq. \eqref{Gscaling}. Red line represents linear dependence. (b) The schematic phase diagram for the rigidity of random spring networks under an internal stress $\sigma_M$.}
\label{Fig2p}
\end{figure}

The non affinity parameter, $\delta \Gamma(\sigma_M,z,\kappa)$, depends on the system's parameters as shown in Fig. \ref{Fig1}c,d. To confirm the prediction of Eq.~\eqref{Gscaling} we plot 
$G-G_0-\frac{5}{6}\sigma_M$ vs. $\sigma_M \delta\Gamma$ and find that all data collapses on to the same curve with a linear dependence, as shown in Fig. \ref{Fig2p}(a).
Interestingly, the scaling prediction in Eq.~\eqref{Gscaling} suggests that $\delta \Gamma$ can be interpreted as a susceptibility of the shear modulus to the internally generated stress. Moreover, $\delta \Gamma$ shows a strong increase close to both rigidity thresholds (Fig. \ref{Fig1} d), implying a large susceptibility to $\sigma_M$ when the system is marginally stable. However, at these stability thresholds $\delta \Gamma$ acquires a strong dependence on $\sigma_M$. This can be understood by considering $\sigma_M$  as an external field that restores rigidity and suppresses the divergence of the strain fluctuations $ \delta\Gamma \sim \left| \Delta z \right|^{-\lambda} $ (for $\sigma_M=0$); this implies a dependence of the form $\delta\Gamma(z_b)\sim(\kappa/\sigma_M)^{-y}$ and   $\delta\Gamma(z_{\rm cf})\sim(\mu/\sigma_M)^{-y'}$ and, taken together with Eq.~\eqref{Gscaling}, explains the origins of the anomalous regimes, where 
\begin{equation}
G(z_b)\sim \kappa ^y\sigma_M^{1-y}, \ \ \ G(z_{\rm cf})\sim \mu ^{y'}\sigma_M^{1-y'}.
\end{equation}
We verified the internal consistency by determining the exponents $y$ and $y'$ at the two stability thresholds from both the scaling of $G$ and $\delta \Gamma$ with $\sigma_M$, as shown in Fig. \ref{Fig2}. 

The schematic phase diagram for the high-motor density limit is shown in Fig.~\ref{Fig2p}(b). Away from the stability thresholds the shear modulus scales linearly with the active stress. This is in contrast with the stiffening behavior of \emph{externally} deformed networks, for which the dependence of the differential elastic modulus goes as the square root of the external stress~ \cite{wyart2008,Sheinman2012Nonlinear}. Thus, there is not necessarily a quantitative correspondence between internally and externally stressed networks, in contrast to suggestions in prior work \cite{Koenderink2009Active}.

\begin{figure}[tp]
\includegraphics[width=\columnwidth]{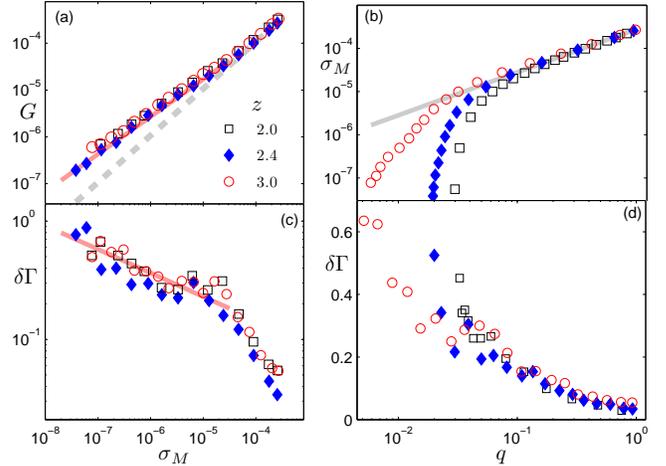}
\caption{The role of inhomogeneity in the motor distribution for different values of $z$ (see legend on panel (a) for every plot in this figure). (a) Shear modulus is presented vs. the normal stress induced by the motors. Solid line represents the power law of $0.8$ while the dashed line contain the mean-filed predictions. (b) The normal stress vs. the motor probability occupation. Solid line represents the mean-field prediction, $ \sigma_M=\sqrt{8}qf $. (c) The differential non-affine fluctuations measure for the same data as in (a) is presented vs. $ \sigma_M $. The solid line represents the power law of $-0.2$. (d) $\delta\Gamma$ vs. $q$. For all the data in this figure $f=10^{-4}$ and $\kappa=10^{-5}$.}
\label{Fig3}
\end{figure}

Finally, we explore the role of inhomogeneity in the distribution of active motors, which shows that critical behavior is not limited to the critical points associated with rigidity percolation. We model inhomogeneous motors by considering the range $q<1$ for different values of $z$ well below the rigidity percolation point, $ z<z_{b} $. In this case the motors only induce a macroscopic stress when the motor density exceeds a $z$-dependent threshold, $q_c\left(z\right) $, as shown in Fig. \ref{Fig3}(b). Concurrent with the development of a macroscopic stress, the network acquires a finite shear rigidity. Near the threshold $q_c$, the motor-induced stress falls significantly below the mean-field prediction ($ \sigma_M=\sqrt{8} q f $) and depends non-linearly on $q$. Interestingly, in this regime ($\sigma_M \ll \sqrt{8} q_c f$) the nonaffine fluctuations become large (see Fig. \ref{Fig3}(d)), diverging with motor stress with an exponent close to $-0.2$, as shown in Fig. \ref{Fig3}(c). Such a divergence, taken together with Eq. \eqref{Gscaling} implies an anomalous, sub-linear scaling of the shear modulus with the motors stress with the exponent $0.8$. Indeed, as shown in Fig. \ref{Fig3}(a), the stiffening of the shear modulus clearly deviates from the mean-field predictions and scales sublinearly with the motor stress with an $0.8$ exponent, consistent with Eq. \eqref{Gscaling}, even when the mean coordination number of the network is well below the rigidity percolation point. Thus, criticality in the form of a divergent susceptibility is characteristic of floppy systems below the rigidity percolation point.

This work demonstrates that motor activity controls the elastic properties of disordered networks by coupling to the differential non-affine fluctuations in the deformation field. This coupling makes elastic deformations more affine and stabilizes the network. Far from the elastic critical points this coupling leads to linear stiffening as a function of the motors stress, as has been observed in several studies of prestressed elastic networks \cite{Budiansky1987Elastic,Stamenovic2000Quantitative}. However, close to the elastic critical points, where the non-affine fluctuations diverge, this coupling leads to anomalous regimes, where the shear modulus scales sub-linearly with the motors stress. Similar stress-stiffening of floppy networks below marginal stability is also found beyond a threshold in the motor density, indicating that a surprising generality of critical fluctuations and divergent susceptibility for systems below the usual rigidity percolation point. 

\begin{acknowledgments}
This work was supported in part by FOM/NWO and in part by a Lewis-Sigler fellowship. The authors thank B. Shklovskii and L. Jawerth for helpful discussions.
\end{acknowledgments}

\appendix
\section{Mean-field approach}
The nonlinear EM approach developed here is based on a scheme to construct
a mapping from the internally stressed lattice network with disordered spring constant, $ \mu_{ij} $,
 with probability density $ P\left( \mu_{ij}\right) $, onto a perfect lattice system
with uniform bond stiffness with the same stress, $\sigma_{M}$. The stress of the network appears due to motors' force $ f $ that acts between each pair of neighboring nodes of the network.
This mapping is realized by an effective uniform central
force interaction, $\mu_{ij}\rightarrow\widetilde{\mu}$. The effective
parameter, $\widetilde{\mu}$$\left(\sigma_{M}\right)$, is determined
by a self-consistency requirement: replacing a random bond in the
uniform EM under stress with a bond drawn from the original probability
density, $P\left(\mu_{ij}\right)$, results in a local fluctuation
in the deformation field, which vanishes when averaged. In addition, we assume that the fluctuations
of the deformations are small compared to the distance between crosslinks.
This approach leads to an integral equation, representing a disorder
average (Eq. (\ref{eq:mEff_Integral})), from which the effective
parameter $\widetilde{\mu}(\sigma_{M})$ can be determined. 
In the following we ignore the contribution of the bending stiffness of the network filaments. Due to this assumption the presented mean-field approach fails in the regime where the elasticity of the network is dominated by the bending stiffness.

\subsection{Effective medium theory}
\label{Effective medium theory} 
We apply the EM theory method to
a network subjected to a uniform internal compression resulting in
a macroscopic isotropic stress $\sigma_{M}$. Similarly to Ref. \cite{Feng85},
we calculate the effective
spring constant using the self-consistency requirement.

The position of a crosslink (network's node) $i$ is given by $\mathbf{R}_{i}=\mathbf{R}_{i}^{0}+\mathbf{u}_{i}$,
where $\mathbf{R}_{i}^{0}$ is the position in the unstressed configuration
and $\mathbf{u}_{i}$ is the displacement field. In order to calculate the elastic constant we apply an infinitesimal external expansional/comressional strain $ \epsilon $ to the network.
The affine displacement due to the applied strain is given by $\mathbf{u}_{i}^{\rm aff}-\mathbf{u}_{j}^{\rm aff}=\epsilon\mathbf{r}_{ij}$,
where $\mathbf{r}_{ij}$ is the vector from $\mathbf{R}^0_{i}$
to $\mathbf{R}^0_{j}$ in the undeformed reference state. Here we allow for non-affine displacements
\begin{equation}
\mathbf{v}_{i}\equiv\mathbf{u}_{i}-\mathbf{u}_{i}^{\rm aff},
\end{equation}
The Hamiltonian of the network is given by
\begin{equation}
H=\frac{1}{2}\sum\limits_{\langle ij \rangle}\mu_{ij} 
\left( \lvert  \mathbf{R}_{ij} \rvert-1 \right)^2  +f\sum\limits_{\langle ij \rangle} \lvert  \mathbf{R}_{ij} \rvert,
\label{Hamiltonian}
\end{equation}
where $ \mathbf{R}_{ij}=\mathbf{R}_{j}-\mathbf{R}_{i} $ and the sums extend over neighboring pairs of vertices.
We assume that the resulting non-affine
relative displacements of neighbouring nodes $i$ and $j$ are much
smaller than the distance between the nodes, 
\begin{equation}
\left|\mathbf{v}_{ij}\right|\equiv\left|\mathbf{v}_{i}-\mathbf{v}_{j}\right|\ll\left|\mathbf{R}_{i}^0-\mathbf{R}^0_{j}\right|=1.
\end{equation}
Thus, we can expand the Hamiltonian around the affine strain configuration
(small $v_{ij}$). Up to second order in $v_{ij}$ and first order in $ \epsilon $ we arrive at~\cite{tang1988percolation,Alexander1998}
\begin{widetext} 
\begin{equation}
H=\underset{\left\langle ij\right\rangle }{\sum}f+\left( f+\epsilon \mu_{ij}\right)  \mathbf{v}_{ij}\cdot\mathbf{r}_{ij}+\frac{1}{2}\left(\mu_{ij}-f\right)\left(\mathbf{v}_{ij}\cdot\mathbf{r}_{ij}\right)^{2}+\frac{1}{2}f\mathbf{v}_{ij}^{2}.\label{eq:nonlinearmodel}
\end{equation}
\end{widetext} 
 The first term represents the expansion/compression energy of the
affine response, while the other terms correspond to the energy difference
due to the non-affine deformation of the stretched/compressed bonds.

The expansion of the whole network corresponds to the global constraint
\begin{equation}
\underset{\left\langle ij\right\rangle }{\sum}\mathbf{v}_{ij}=0.\label{eq:constraint}
\end{equation}

To investigate the elastic behavior of the model in Eq. (\ref{eq:nonlinearmodel}),
we set up an effective medium theory. In the EM approach, we mimic
the disordered system by the regular one with an effective parameter,
i.e. $\mu_{ij}\rightarrow\widetilde{\mu}$ and the same stress $\sigma_{M}$.
In other words, the EM network may be globally expanded by applying
the force that assures mechanical equilibrium for the affine, $\mathbf{v}_{ij}=0$,
configuration. Thus, the EM system has the Hamiltonian, given by 
\begin{widetext} \begin{equation}
H_{EM}=\underset{\left\langle ij\right\rangle }{\sum}f+\left( f+\epsilon \widetilde{\mu}\right) \mathbf{v}_{ij}\cdot\mathbf{r}_{ij}+\frac{1}{2}\left(\widetilde{\mu}-f\right)\left(\mathbf{v}_{ij}\cdot\mathbf{r}_{ij}\right)^{2}+\frac{1}{2}f\mathbf{v}_{ij}^{2}+\mathbf{\Lambda}_{ij}\cdot\mathbf{v}_{ij}\label{eq:H_EM}
\end{equation}\end{widetext}
 where $\mathbf{\Lambda}_{ij}=-\left( f+\epsilon\widetilde{\mu}\right) \mathbf{r}_{ij}$.
To calculate the effective parameter $\widetilde{\mu}$ we demand
self-consistency of the EM~\cite{Feng85}. The self-consistency requirement
in this context can be formulated as follows: the non-affine displacement
induced by the replacement of a single bond in the EM vanishes on
average, 
\begin{equation}
\left\langle \mathbf{v}_{nm}\right\rangle =0.\label{eq:SelfConsistence}
\end{equation}
 Here, the average is taken over the distribution of the $nm$ bond
in the original disordered system, i.e. according to the probability
density $P\left(\mu_{nm}\right)$. To calculate the displacement $\mathbf{v}_{nm}$
after the replacement we solve the perturbed EM Hamiltonian that is
given by 
\begin{equation}
H_{EM}+\frac{1}{2}\left(\mu_{nm}-\widetilde{\mu}\right)\left(\mathbf{v}_{nm}\cdot\mathbf{r}_{nm}\right)^{2}+\mathbf{v}_{nm}\cdot\mathbf{r}_{nm}\epsilon\left(\mu_{nm}-\widetilde{\mu}\right)
\end{equation}
 In the configuration that minimizes the energy, the displacement
of the $nm$ bond is given by 
\begin{equation}
\mathbf{v}_{nm}=\frac{\mathbf{r}_{nm}\epsilon\left(\mu_{nm}-\widetilde{\mu}\right)}{\mu_{EM}+\mu_{nm}-\widetilde{\mu}},\label{eq:DisplacementMainText}
\end{equation}
 where $\mu_{EM}$ is the displacement of the $nm$ bond in the \emph{unperturbed}
EM network due to a unit force acting along the $nm$ bond. 

\subsubsection{The calculation of $\widetilde{\mu}_{EM}$\label{AppA}}

In this Section we calculate $\mu_{EM}$---the displacement of the
$nm$ bond in the \emph{unperturbed} EM network due to a unit force
$\mathbf{r}_{nm}$ acting on the $nm$ bond.

The dynamical matrix of the unperturbed EM Hamiltonian (\ref{eq:H_EM})
is given by 
\begin{equation}
D_{ij}=\begin{cases}
-\left(\widetilde{\mu}-f\right)\mathbf{r}_{ij}\otimes\mathbf{r}_{ij}+f\mathbb{I} & i\neq j\\
\underset{j\neq i}{\sum}\left[\left(\widetilde{\mu}-f\right)\mathbf{r}_{ij}\otimes\mathbf{r}_{ij}+f\mathbb{I}\right] & i=j
\end{cases},
\end{equation}
 where $\mathbb{I}$ is the unit tensor and $\otimes$ is the external
product. The Fourier transform of $D$ is given by 
\begin{eqnarray}
D\left(\mathbf{k}\right) & = & \underset{ij}{\sum}D_{ij}e^{i\mathbf{k}\cdot\mathbf{r}_{ij}}=\nonumber \\
 & = & \underset{\mathbf{r}}{\sum}\left(\left(\widetilde{\mu}-f\right)\mathbf{r}\otimes\mathbf{r}+f\mathbb{I}\right)\left(1-e^{i\mathbf{k}\cdot\mathbf{r}}\right)
\end{eqnarray}
 where $\mathbf{r}$ runs over all unit bond vectors. The unit force
acting on the $nm$ bond is given by 
\begin{equation}
\mathbf{f}_{i}=\mathbf{r}_{nm}\left(\delta_{i,n}-\delta_{i,m}\right),
\end{equation}
 so that its Fourier transform is 
\begin{equation}
\mathbf{f}\left(\mathbf{k}\right)=\underset{i}{\sum}\mathbf{f}_{i}e^{i\mathbf{k}\cdot\mathbf{R}_{i}}=\mathbf{r}_{nm}\left(1-e^{i\mathbf{k}\cdot\mathbf{r}_{nm}}\right).
\end{equation}
 Thus the Fourier transform of the displacement field is given by
\begin{equation}
\mathbf{v}\left(\mathbf{k}\right)=-D^{-1}\left(\mathbf{k}\right)\cdot\mathbf{f}\left(\mathbf{k}\right).
\end{equation}
 The displacement of the $nm$ bond due to the unit force is
\begin{widetext} 
\begin{alignat*}{1}
\mu_{EM}^{-1} & =\frac{1}{N}\mathbf{r}_{nm}\cdot\underset{\mathbf{k}}{\sum}\mathbf{v}\left(\mathbf{k}\right)\left(e^{-i\mathbf{k}\cdot\mathbf{r}_{nm}}-1\right)=-\underset{\mathbf{k}}{\sum}\mathbf{r}_{nm}\cdot\mathbf{f}\left(\mathbf{k}\right)D^{-1}\left(\mathbf{k}\right)\left(e^{-i\mathbf{k}\cdot\mathbf{r}_{nm}}-1\right)\\
=\widetilde{\mu}^{-1} & \frac{2d}{\mathcal{Z}}\left[1-\frac{f}{dN\left(\widetilde{\mu}-f\right)}\underset{\mathbf{k}}{\sum}Tr\left\{ \frac{\underset{\mathbf{r}}{\sum}\left(1-e^{i\mathbf{k}\cdot\mathbf{r}}\right)}{\underset{\mathbf{r}}{\sum}\left(\mathbf{r}\otimes\mathbf{r}+\frac{f}{\widetilde{\mu}-f}\mathbb{I}\right)\left(1-e^{i\mathbf{k}\cdot\mathbf{r}}\right)}\right\} \right].
\end{alignat*}\end{widetext} 
 For a highly coordinated lattice the sum over $\mathbf{r}$ may be
well approximated by the integral over the sphere that includes all
the neighbouring crosslinks and, since the sum over $\mathbf{k}$
is dominated by the small $\mathbf{k}\cdot\mathbf{r\ll1}$ values,
$\mu_{EM}$ may be approximated by
\begin{widetext} 
\begin{eqnarray*}
\mu_{EM}^{-1} & \simeq & \widetilde{\mu}^{-1}\frac{2d}{\mathcal{Z}}\left[1-\frac{f}{dN\left(\widetilde{\mu}-f\right)}Tr\left\{ \frac{\oint\left(\mathbf{k}\cdot\mathbf{r}\right)^{2}d^{d-1}\mathbf{r}}{\oint\left(\mathbf{k}\cdot\mathbf{r}\right)^{2}d^{d-1}\mathbf{r}\left(\mathbf{r}\otimes\mathbf{r}+\frac{f}{\widetilde{\mu}-f}\mathbb{I}\right)}\right\} \right]\\
 & = & \widetilde{\mu}^{-1}\frac{2d}{\mathcal{Z}}\left[1-\frac{f}{d\left(\widetilde{\mu}-f\right)}\left(\frac{1}{\frac{3}{2+d}+\frac{f}{\widetilde{\mu}-f}}+\frac{d-1}{\frac{1}{2+d}+\frac{f}{\widetilde{\mu}-f}}\right)\right].
 \label{muEM}
\end{eqnarray*}
\end{widetext}

\subsubsection{An effective elastic constant}
Given Eqs. (\ref{eq:SelfConsistence},\ref{eq:DisplacementMainText},\ref{muEM}),
the self-consistency Eq. (\ref{eq:SelfConsistence}) leads to the
following equation for the effective parameter%
\footnote{In the unstressed regime, $ f\rightarrow 0 $, Eq. (\ref{eq:mEff_Integral}) reduces to Eq.
(9) in Ref. \cite{Feng85}.}
\begin{equation}
\int_{0}^{\infty}\frac{\mu_{ij}-\widetilde{\mu}\left(\sigma_{M}\right)}{\mu_{EM}+\mu_{ij}-\widetilde{\mu}\left(\sigma_{M}\right)}P\left(\mu_{ij}\right)d\mu_{ij}=0.\label{eq:mEff_Integral}
\end{equation}

The approach presented in this section allows one to calculate the
elastic parameters of a system with a given topology and elastic constant
distribution in the nonlinear elastic regime.
Eq. (\ref{eq:mEff_Integral}) may be solved numerically for any realization
of the spring constant probability density, $P\left(\mu_{ij}\right)$.
Knowing the effective spring constant, $ \widetilde{\mu} $, one obtains all the elastic constants of the network and the relation between the motors' applying force, $ f $ and the global normal network's stress, $ \sigma_M $.
In the next section we demonstrate the presented
method using the particular example of diluted regular networks when
Eq. (\ref{eq:mEff_Integral}) can be solved analytically.

\subsection{Diluted regular networks}
\label{Diluted regular networks} 


In this Section we use the mean-field
solution presented above using for particular example of bond-diluted
regular networks. 
The probability density for the spring constants
for such a network is given by
\begin{equation}
P\left(\mu_{ij}\right)=\frac{z}{\mathcal{Z}}\delta\left(\mu_{ij}-\mu\right)+\left(1-\frac{z}{\mathcal{Z}}\right)\delta\left(\mu_{ij}\right).
\end{equation}
Networks of this
kind are referred to as diluted spring networks or the central-force
elastic percolation model. The linear elastic response of diluted,
unstressed lattices has been extensively studied \cite{feng1984,Feng85}.
Here we show how these results generalize for internally stressed
networks.

In this case the Eq. \eqref{eq:mEff_Integral} becomes
\begin{widetext} 
\begin{equation}
\frac{\frac{\widetilde{\mu}}{\mu-\widetilde{\mu}}+\frac{2d}{\mathcal{Z}}\left[1-\frac{f}{d\left(\widetilde{\mu}-f\right)}\left(\frac{1}{\frac{3}{2+d}+\frac{f}{\widetilde{\mu}-f}}+\frac{d-1}{\frac{1}{2+d}+\frac{f}{\widetilde{\mu}-f}}\right)\right]}{1-\frac{2d}{\mathcal{Z}}\left[1-\frac{f}{d\left(\widetilde{\mu}-f\right)}\left(\frac{1}{\frac{3}{2+d}+\frac{f}{\widetilde{\mu}-f}}+\frac{d-1}{\frac{1}{2+d}+\frac{f}{\widetilde{\mu}-f}}\right)\right]}=\frac{z/\mathcal{Z}}{1-z/\mathcal{Z}}.
\end{equation}
\end{widetext} 
The solution of this equation provides the spring constant of the
effective medium, $\widetilde{\mu}$, and, therefore, knowing the
geometry of the original lattice one can easily calculate all the
elastic constants of the network.

\bibliographystyle{apsrev}
\bibliography{NonLinear} 

\end{document}